\documentclass[twocolumn,aps,prb,longbibliography]{revtex4-2}
\usepackage{graphicx}
\usepackage{amsfonts}
\usepackage{amsmath}
\usepackage{amssymb}
\usepackage{bm}
\usepackage{hyperref}
\usepackage{slashed}
\usepackage{braket}

\usepackage{color}

\begin{document}

\title{Chiral and nodal superconductors in t-J model with valley contrasting flux on  triangular moir\'e lattice}
\author{Boran Zhou}
\author{Ya-Hui Zhang}
\affiliation{Department of Physics and Astronomy, Johns Hopkins University, Baltimore, Maryland 21218, USA}

\begin{abstract}

Recent experimental progresses have made it possible to simulate spin 1/2 Hubbard model on triangular lattice in moir\'e materials formed by transition metal dichalcogenide (TMD) heterobilayer or homobilayer. In twisted TMD homobilayer, a vertical electric field can induce a valley contrasting flux in the hopping term. In this paper we study possible superconductors from a t-J model with valley contrasting flux $\Phi$ using the slave boson mean field theory. We obtain a phase diagram with doping $x$ and $\Phi$. A finite $\Phi$ breaks spin rotation symmetry and the pairing symmetry is a superposition of spin singlet $d-id$ and spin triplet $p+ip$. There are two topological phase transitions when tuning $\Phi$ from $0$ to $\pi$, with three Dirac nodes at one transition and one single Dirac node at the other transition. We also discuss the effects of van Hove singularity and a three-site correlated hopping term on the pairing strength. Lastly, we demonstrate that a small anisotropy term breaking the $C_3$ rotation can lead to a time reversal invariant nodal superconductor  connected to the $d_{x^2-y^2}$ superconductor on square lattice.

\end{abstract}

\maketitle

\emph{Introduction}-- Superconductivity from doping a Mott insulator has been intensively studied in the last several decades after the discovery of the high Tc superconductor in cuprates\cite{RevModPhys.78.17}. Recently, moir\'e superlattices emerge to be a wonderful platform to simulate strongly correlated physics\cite{cao2018correlated,Wang2019Signatures,Wang2019Evidence,yankowitz2019tuning,Wang2019Signatures,chen2019tunable,lu2019superconductors,Cao2019Electric,Liu2019Spin,Shen2019observation,polshyn2020nonvolatile,chen2020electrically,sharpe2019emergent,serlin2020intrinsic,tang2020simulation,regan2019optical,wang2019magic,gu2021dipolar,zhang2021correlated,li2021continuous,ghiotto2021quantum,li2021quantum}. Superconductivity has been experimentally reported in twisted bilayer graphene and twisted multilayer graphene system with alternating twist angles\cite{cao2018unconventional,yankowitz2019tuning,park2021tunable,hao2021electric}. However, mechanism of the superconductivity there is still under debate. Theoretical study in TBG is hard due to the lack of a simple lattice model which is obstructed by the fragile topology\cite{po2018origin}.  On the other hand, moir\'e superlattice based on transition metal dichalcogenide (TMD) is believed to be described by a simple Hubbard model\cite{wu2018hubbard} similar to that of the cuprate. This offers an opportunity to study superconductivity in Hubbard model or  t-J model on triangular lattice. 

Moir\'e superlattice  can be formed by twisting TMD homo-bilayer or hetero-bilayer. In both cases the resulting Hubbard model has two flavors from the two valley, which are locked to the spins due to the strong Ising spin-orbit coupling of the valence band.  In the hetero-bilayer case, it was shown that there is a good SU(2) spin rotation symmetry in the valley (spin) space\cite{wu2018hubbard}. Thus the physics of  doping the Mott insulator at $\nu_T=1$ is captured by the standard t-J model. In contrast, for the twisting TMD homo-bilayer \footnote{Here we focus on the AA stacking. The case of AB stacking is very different\cite{zhang20214}.}, the low energy model has a valley contrasting flux $\Phi$ induced by a displacement field in the $z$ direction\cite{pan2020band,wang2021staggered}. The valley contrasting flux origins from the inversion symmetry breaking within each valley and is known to exist also in graphene moir\'e systems\cite{zhang2019bridging}. In such a case, the SU(2) spin rotation symmetry is broken down to U(1) except at $\Phi=0$ mod $2\pi$. The t-J model with $\Phi=0$ has been found to host a $d-id$ superconductor by the slave boson mean field theory\cite{wang2004doped}. But the fate at generic $\Phi$ is unknown, although there are already a few studies at weak coupling limit \cite{wu2022pair,wu2022pair2,wietek2022tunable}. For application to twisted TMD homobilayer, it is desirable to obtain a phase diagram of $(n,\Phi)$ given that both the density $n$ and $\Phi$ can be conveniently tuned in the dual gated sample.

In this paper we perform a slave boson mean field study of the t-J model with a generic valley contrasting flux $\Phi$.  The flux $\Phi$ enters both the hopping term and the $J$ term through a Dzyaloshinskii-Moriya (DM) interaction. At finite $\Phi$, we find that the $d-id$ spin-singlet pairing is mixed with a $p+ip$ spin-triplet pairing. By changing $\Phi$ at fixed density $n$, we find two topological phase transitions between Chern number $|C|=2$ and $|C|=1$ through three Dirac nodes and one Dirac node respectively. We also notice the correlations between the density of states (DOS) and the pairing strength. Pairing is stronger at the van hove singularity. Chiral superconductivity has been observed in numerical simulations in the SU(2) symmetric limit\cite{jiang2020topological,huang2022emergent,huang2022topological}, our study demonstrates a new route to tune a topological phase  transition through changing the valley contrasting flux.  We also study the effect of strain which breaks the $C_3$ rotation symmetry. We find that a small strain favors a nodal superconductor with the same  symmetry as the $d_{x^2-y^2}$ pairing in cuprates. This suggests that a nematic nodal superconductor is a strong competing state. Indeed recent numerical simulation shows that adding a $J_2$ term can lead to a transition from the chiral superconductor to the nematic nodal superconductor\cite{huang2022topological}.

\begin{figure}[h]
    \centering
    \includegraphics[width=\linewidth]{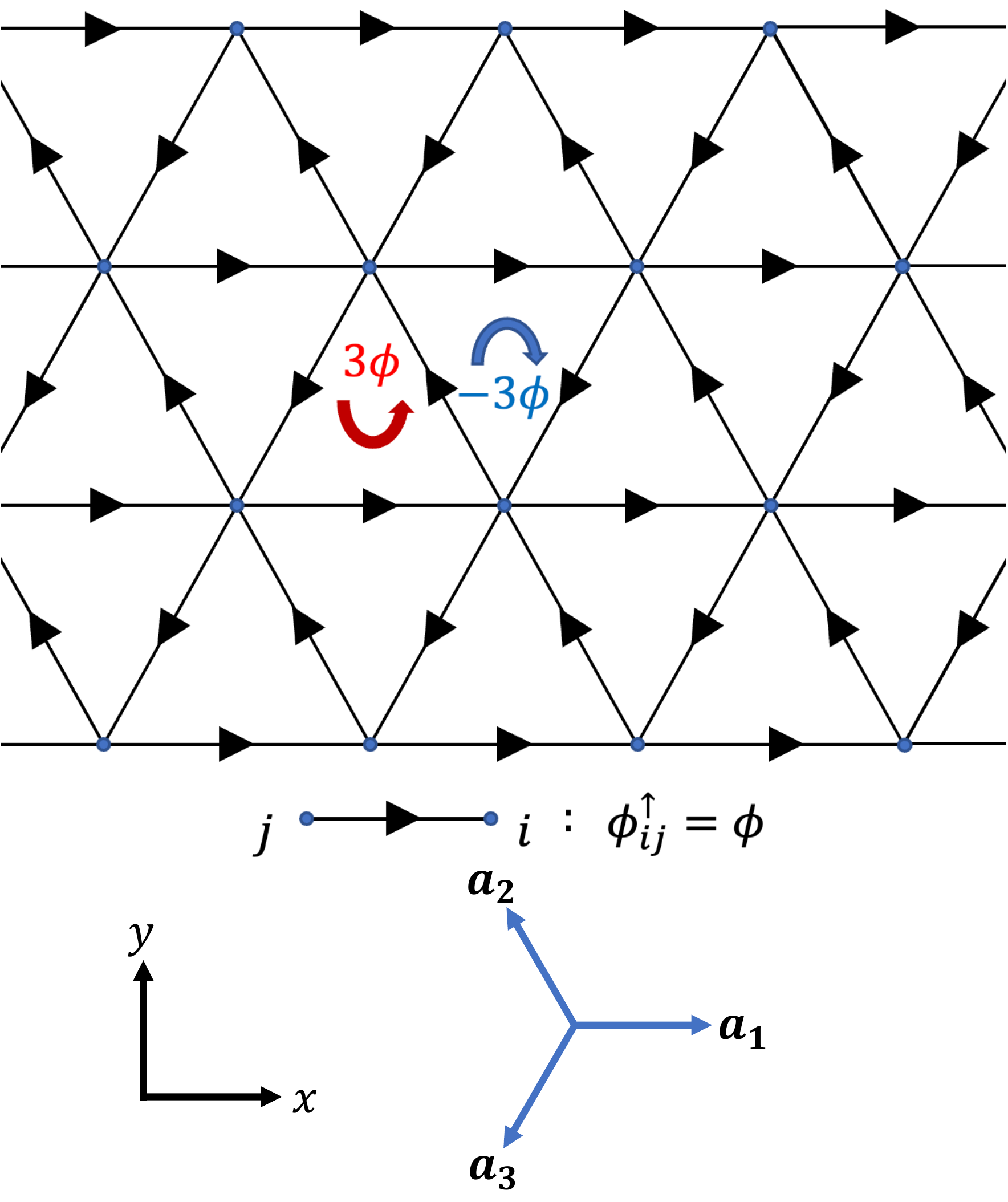}
    \caption{(color online). Illustration of the flux pattern for the valley $\uparrow$. The other valley is related by time reversal symmetry and has opposite flux pattern.}
    \label{Fig1}
\end{figure}

\emph{Model}-- The twisted TMD homobilayer can be described by a generalized triangular-lattice Hubbard model \cite{PhysRevResearch.2.033087}:
\begin{equation}\label{hubbard}
\begin{split}
    H&=-\sum_{\braket{ij},s}\left(te^{i\phi_{ij}^s} c^\dagger_{i,s}c_{j,s}+\text{H.c.}\right)\\
    &+U\sum_{i}n_{i\uparrow}n_{i\downarrow}-\mu\sum_i(n_{i\uparrow}+n_{i\downarrow}),
\end{split}
\end{equation}
where  $s=\uparrow,\downarrow$ labels the valley index which is locked to the spin.   $\braket{ij}$ represents the nearest-neighbor (NN) bond. We have $\phi_{ij}^{\uparrow}=-\phi_{ij}^{\downarrow}$ due to the time reversal symmetry. As illustrated in Fig.\ref{Fig1}, $\phi_{ij}^\uparrow=\pm\phi$ depending on the direction of the bond. There is a valley contrasting flux $\Phi=3\phi$ and $-\Phi$ in the two types of triangles for one valley.  We label $\sigma_a$ as Pauli matrices in the valley space. We have time reversal symmetry $\mathcal{T}$ acting as $\sigma_x \mathcal K$, where $\mathcal K$ is the complex conjugate. The mirror reflection operators $\mathcal{M}_x$ and $\mathcal{M}_y$ act as $c_s(x,y)\rightarrow c_{s}(x,-y)$ and $c_s(x,y)\rightarrow (\sigma_x)_{ss'}c_{s'}(-x,y)$ respectively.   $C_6$ around site $\mathbf{x}$ acts as: $c_{s}(\mathbf x)\rightarrow (\sigma_x)_{s s'} c_{s'}(C_6 \mathbf x)$.  Two basis vectors of the lattice are $\mathbf{a_1}=(1,0)$ and $\mathbf{a_2}=(-\frac{1}{2},\frac{\sqrt{3}}{2})$. The corresponding reciprocal basis vectors are $\mathbf{b_1}=(2\pi,\frac{2\pi}{\sqrt{3}})$ and $\mathbf{b_2}=(0,\frac{4\pi}{\sqrt{3}})$.  $\phi$ and $\phi+\frac{2\pi}{3}$ are gauge equivalent upon a transformation: $c_s(\mathbf{x}) \rightarrow (e^{i\frac{\mathbf{x}}{3}\cdot(\mathbf{b_1}+\mathbf{b_2})\sigma_z})_{ss'}c_{s'}(\mathbf{x}) $.   We can combine a particle hole transformation with the gauge transformation: $c_s(\mathbf{x}) \rightarrow (e^{i\frac{\mathbf{x}}{6}\cdot(\mathbf{b_1}+\mathbf{b_2})\sigma_z}\sigma_x)_{ss'} c_{s'}^\dagger(\mathbf{x})$, which maps $\Phi$ to $\Phi+\pi$ and the density $n$ to $2-n$ \cite{wu2022pair}. 

In the strong coupling limit $t\ll U$ and filling factor $n<1$, we can use the standard $t/U$ expansion  \cite{hubavc2010brillouin} to obtain a t-J model \cite{PhysRevResearch.2.033087}:
\begin{equation}\label{eqtJ}
\begin{split}
    H&=-\sum_{\braket{ij},s}P\left(te^{i\phi_{ij}^s} c^\dagger_{i,s}c_{j,s}+\text{H.c.}\right)P+J\sum_{\braket{ij}}\Big[S_i^zS_j^z\\
    &+\cos(2\phi^\uparrow_{ij})\sum_{\alpha=x,y}S_i^\alpha S_j^\alpha+\sin(2\phi_{ij}^\uparrow)\left(S_i^xS_j^y-S_i^yS_j^x\right)
     \\
    &-\frac{1}{4}n_in_j\Big]-\mu\sum_{i}(n_{i\uparrow}+n_{i\downarrow}),
\end{split}
\end{equation}
where $J=\frac{4t^2}{U}$ and $P$ is the projection operator which forbids double occupancy. For $\phi\ne0$, the system only has a U(1) spin rotation symmetry generated by $\sigma_z$. At $n=1$, the ground state is an XY ferromagnetic phase for $|\phi|\in (\frac{\pi}{3},\frac{2\pi}{3})$, $120^\circ$ $\text{AF}^{+}$ phase for $\phi\in(0,\frac{\pi}{3})\cup(\pi,\frac{4\pi}{3})$ and $120^\circ$ $\text{AF}^-$ phase for $\phi\in(\frac{2\pi}{3},\pi)\cup(\frac{5\pi}{3},2\pi)$  \cite{PhysRevResearch.2.033087}. 
The ground state remains unexplored at finite doping with $n=1-x$. We expect that the magnetic order is suppressed  by the doping \cite{wang2004doped}. The major focus of this paper is to explore the possibility of superconducting phase in the small doping regimes.  For the $n>1$ case, we can apply the particle-hole transformation to map the model to the $n<1$ model above, but with a change of the flux $\phi \rightarrow \phi+\pi$.

\emph{Slave boson mean field theory}-- Since it is difficult to handle the projection operator directly, we use the U(1) slave boson theory \cite{RevModPhys.78.17,PhysRevB.65.014502} to deal with the model. We focus our discussion in the region $n<1$, while the $n>1$ regime can be obtained by the particle hole transformation.  The electron operator can be represented as $c^\dagger_{i,s}=f^\dagger_{i,s}b_i$. By decoupling the interaction part in hopping and pairing channel, we use the following mean field Hamiltonian to perform the calculation:
\begin{equation}
\label{MF}
\begin{split}
H_{\text{MF}}&=-\sum_{\braket{ij},s}(te^{i\phi^s_{ij}}\braket{b_{i}b^{\dagger}_{j}}f^\dagger_{i,s}f_{j,s}+\text{H.c.})\\
&-\frac{J}{8}\sum_{\braket{ij},s}\left[\left(\chi^*_{ij,s}+2e^{i2\phi^s_{ij}}\chi^*_{ij,-s}\right)f^\dagger_{i,s}f_{j,s}+\text{H.c.}\right]\\
&+\frac{J}{8}\sum_{\braket{ij}}\left[\left(\Delta_{ij}+2\Delta_{ji}e^{i2\phi^\uparrow_{ij}}\right)f^\dagger_{i,\uparrow}f^\dagger_{j,\downarrow}+(i\leftrightarrow j)+\text{H.c.}\right]\\
&-\mu\sum_i(n_{i\uparrow}+n_{i\downarrow}),
\end{split}
\end{equation}
with $\chi_{ij,s}=2\braket{f^\dagger_{i,s}f_{j,s}}$ and $\Delta_{ij}=2\braket{f_{i\uparrow}f_{j\downarrow}}$. The chemical potential $\mu$ is tuned to make $\frac{1}{N}\sum_{i,s}\braket{n_{i,s}}=1-x$, where $x$ is the doping level. We ignore the magnetic order since it is suppressed on doping level greater than a critical value $x_c$ \cite{wang2004doped}.  Moreover, we assume that the system respects the translation symmetry, then the bosons are condensed and satisfying $\braket{b_ib^\dagger_j}=\braket{b^\dagger_jb_i}=x$ \cite{wang2004doped}. We can simply take $\langle b \rangle=\sqrt{x}$.

\begin{figure}[h]
    \centering
    \includegraphics[width=\linewidth]{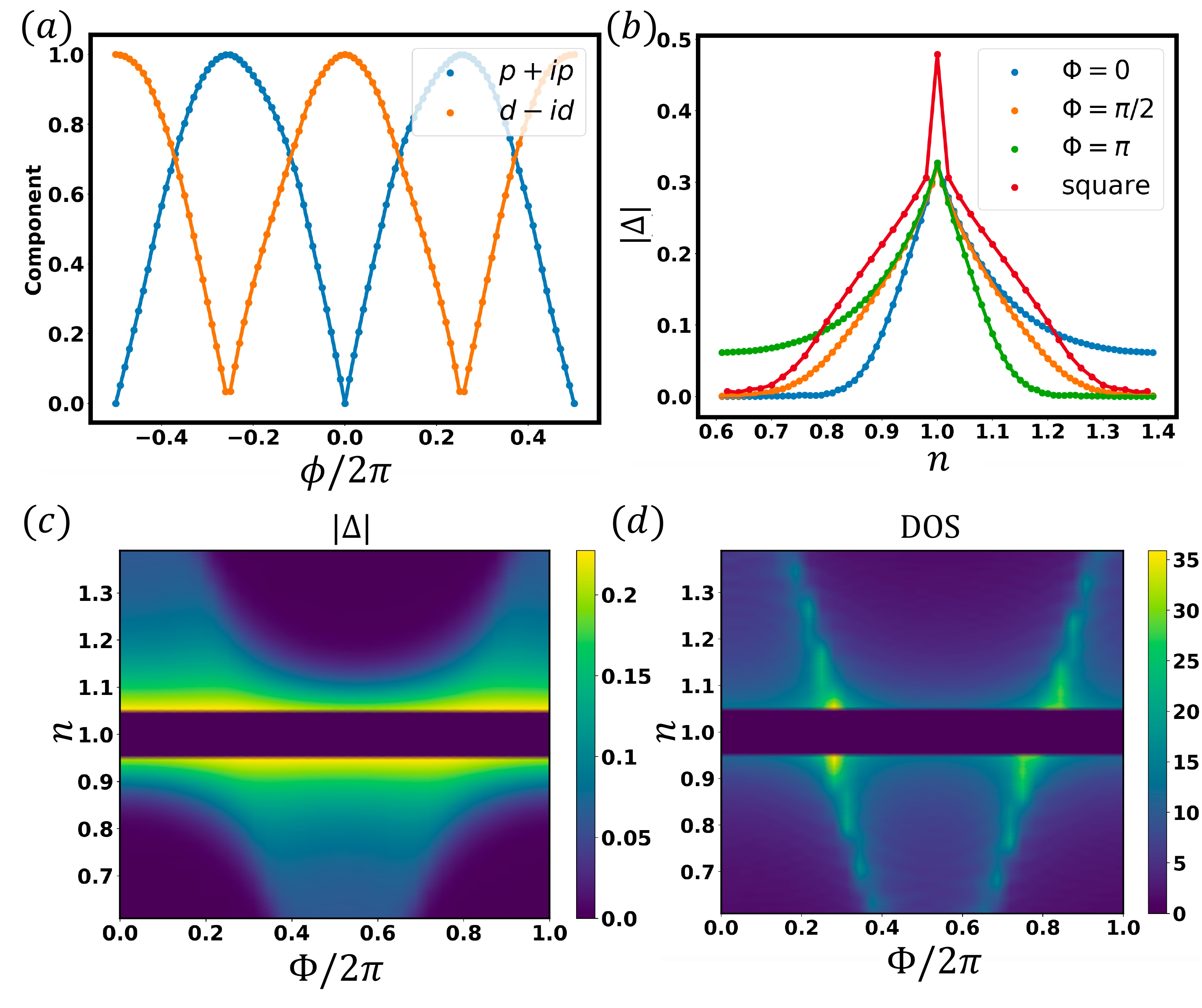}
    \caption{(color online). $t/J=3$. (a) Dependence of $p+ip$ and $d-id$ pairing's component on $\phi$ at $n=0.9$. (b) Dependence of $|\Delta|=|\Delta_{ij}|$ on the density $n$ while flux $\Phi=0,\frac{\pi}{2}$ on triangular lattice and $\Phi=0$ on square lattice. (c) Dependence of $|\Delta|$ on $\Phi$ and $n$. (d) Dependence of DOS on $\Phi$ and $n$. In (c) and (d), the region $0.95<n<1.05$ is removed because the superconductivity is expected to be suppressed because the slave boson condensation $\langle b \rangle =\sqrt{x}$ is weak in this regime. }
    \label{Fig2}
\end{figure}

As for $n>1$ case, we can apply a particle hole transformation $c_{i,s}\rightarrow c^\dagger_{i,-s}$ on the original Hamiltonian and apply the same procedure. By solving the self-consistent equation, we obtained the order parameter dependence on $\phi$ and $n$. The result is shown in Fig.\ref{Fig2}. 

\emph{P wave vs d wave}-- The special case with $\Phi=0$ has been studied before \cite{wang2004doped}. There   $\Delta_{ij}=\Delta e^{\pm i2\theta_{ij}}$ is in the  $d_{x^2-y^2}\pm id_{xy}$ pairing symmetry. With a finite $\Phi$, the spin rotation symmetry is broken down to U(1) generated by the $S_z$ rotation. Therefore, the relation $\Delta_{ij}=\Delta_{ji}$ would not hold anymore and generically the pairing symmetry is a superposition of spin-triplet p wave and spin-singlet d wave.  $\Delta_{ij}$ and $\Delta_{ji}$ have the same magnitude but different phases. Since $\Delta_{ij}$ respects $C_3$ symmetry instead of $C_6$ symmetry, the angular momentum is defined mod $3$. The $p\pm ip$ and $d\mp id$ pairings are mixed, we can define their components as $\Delta_p=\frac{1}{2|\Delta_{ij}|}|\Delta_{ij}-\Delta_{ji}|$ and $\Delta_d=\frac{1}{2|\Delta_{ij}|}|\Delta_{ij}+\Delta_{ji}|$.   The components' dependence on $\phi$ is shown in Fig.\ref{Fig2}(a). 

We note that the superconducting order breaks the time reversal symmetry $\mathcal{T}$ and mirror reflection symmetry $\mathcal{M}_x(\mathcal{M}_y)$, but satisfied the combined mirror time reversal $\mathcal{M}_x\mathcal{T}(\mathcal{M}_y\mathcal{T})$ symmetry combined with a U(1)  transformation.

\begin{figure}[h]
    \centering
    \includegraphics[width=\linewidth]{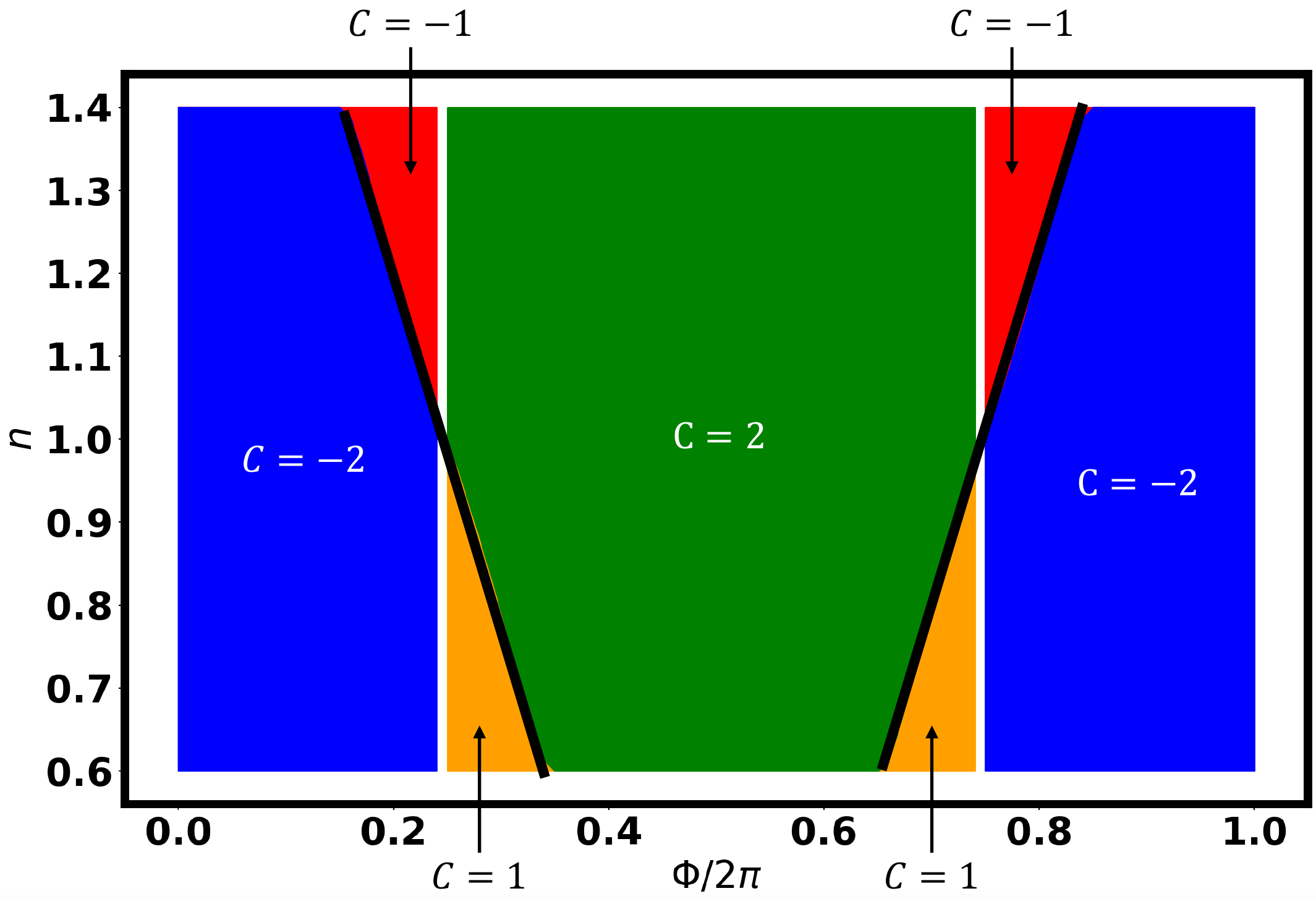}
    \caption{(color online). $t/J=3$. Dependence of Chern number on $n$ and $\Phi$ for fixed chirality. The black and white line corresponds to one and three Dirac nodes in the dispersion plot.}
    \label{Fig3}
\end{figure}

\emph{Pairing strength}-- We study the dependence of $|\Delta_{ij}|$ on different value of $n$ and $\Phi$.  As shown in Fig.\ref{Fig2}(b)(c), the superconducting gap $|\Delta|$ is larger at electron doping than hole doping when $0\le\Phi<\frac{\pi}{2}$ and $\frac{3\pi}{2}<\Phi\le2\pi$, and smaller when $\frac{\pi}{2}<\Phi<\frac{3\pi}{2}$. The reason is that the DOS of the free dispersion does not respect particle hole symmetry, and the Van Hove peak locates at $n>1$ for the former case, $n<1$ for the latter case.   The apparent correlation between the pairing strength and the DOS in Fig.~\ref{Fig3}(d) suggest that we should look for superconductor close to the Van Hove singularity, which is tunable by gating in the twisted TMD homo-bilayer.

\emph{Topological property}-- The chiral superconductor is known to host a non-zero Chern number, which could be calculated via \cite{PhysRevB.61.10267}:
\begin{equation}
    C=\frac{1}{4\pi}\int_{\text{BZ}}d^2\mathbf{k}\left[\hat{\textbf{m}}\cdot\left(\partial_{k_x}\hat{\textbf{m}}\times\partial_{k_y}\hat{\textbf{m}}\right)\right],
\end{equation}
where $\hat{\textbf{m}}=\frac{1}{\sqrt{\epsilon_k^2+|\Delta_k|^2}}(\text{Re}\Delta_k, \text{Im}\Delta_k, \epsilon_k)$. The definition is equivalent to the winding number of $\Delta_k$ as $k$ moves around the Fermi surface in the anti-clockwise direction. We can also calculate the winding number by counting the number of $\Delta$'s zero points inside the Fermi sphere, the formula is $C=\#_1-\#_2$, where $\#_1$ and $\#_2$ represent the number of saddles and sources respectively. The phase diagram is shown in Fig.\ref{Fig3} provided that the chirality is fixed. The change in Chern number equals to the number of Dirac nodes at the transition point. There are two kinds of transitions, which corresponds to one Dirac node (at $\kappa$ point) and three Dirac nodes (inside the mini Brillouin zone) closing the gap respectively. For details, please see Appendix.\ref{appendixA}. 

\begin{figure}[h]
    \centering
    \includegraphics[width=\linewidth]{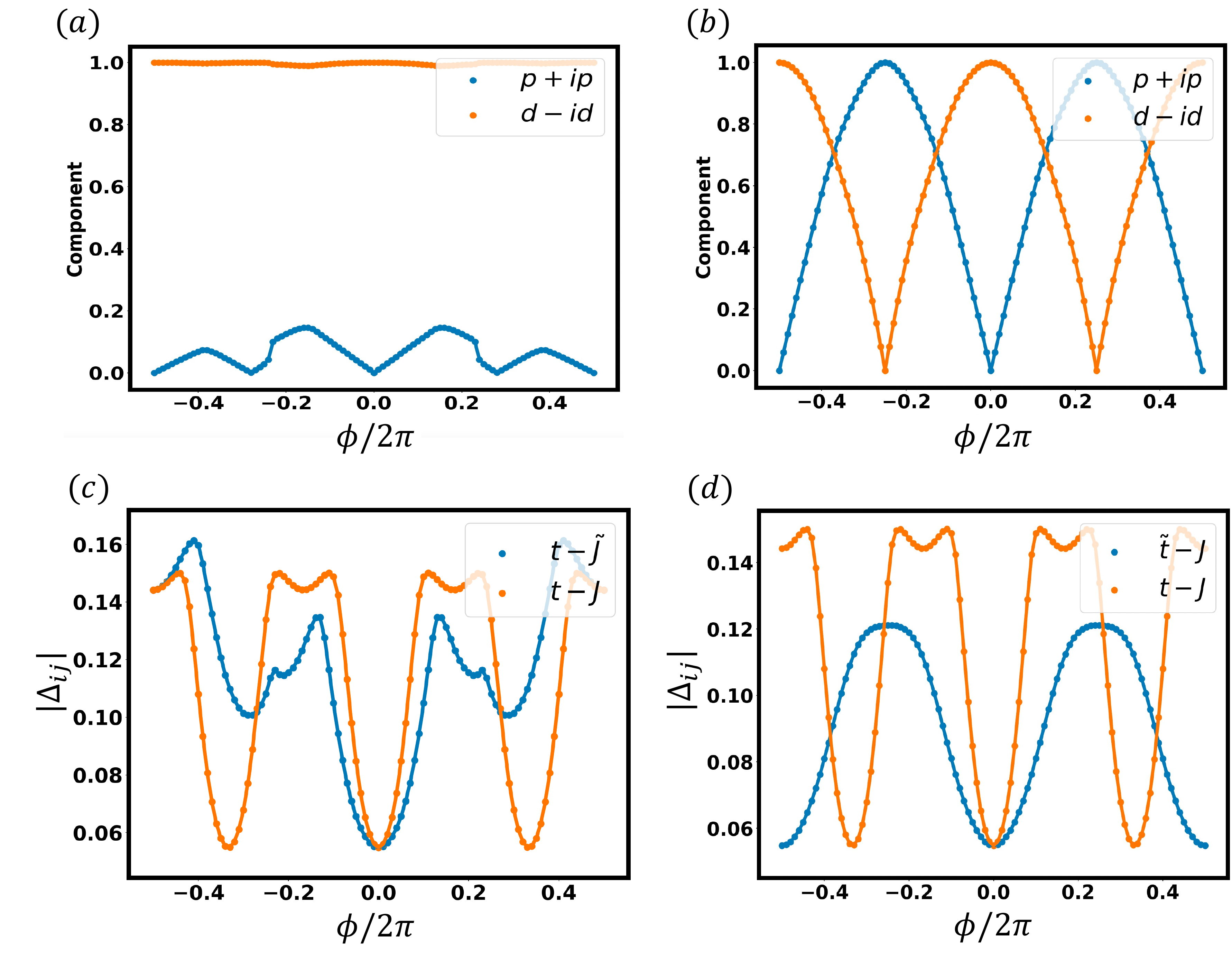}
    \caption{(color online). $t/J=3$. (a)-(b) Components of $p+ip$ and $d-id$ at $n=0.9$ for t-$\tilde{\text{J}}$ model and $\tilde{\text{t}}$-J model respectively. (c)-(d) The superconducting gap $|\Delta_{ij}|$ at $n=0.88$ for t-$\tilde{\text{J}}$ model and $\tilde{\text{t}}$-J model respectively. }
    \label{Fig4}
\end{figure}

\emph{Effects of $\phi$}-- The phase of hopping parameters $\phi_{ij}^s$ plays two roles in our model: (1) It can produce an effective magnetic flux in the $t$ term. (2) It enters the spin-spin coupling $J$ term. The first one changes the distribution of DOS and the second one changes the form of pairing. To study these two effects in detail, we change the kinetic term and the interaction term in Eq.\ref{eqtJ} to the conventional t-J model's form, which are named $\tilde{\text{t}}$-J model and t-$\tilde{\text{J}}$ model respectively. We note that $\phi$ and $\phi+\frac{2\pi}{3}$ could no more be related via a gauge transformation here. In t-$\tilde{\text{J}}$ model, we find that the superconducting order parameter returns to nearly pure $d$ wave pairing shown in Fig.\ref{Fig4}(a), but the hopping term still acquires a phase based on the mean field calculation. As shown in Fig.\ref{Fig4}(b),  the components' dependence on $\phi$ in $\tilde{\text{t}}$-J model is nearly the same as the original model. Therefore, we can conclude that the pairing symmetry is decided by the $J$ term as expected.  In $\tilde{\text{t}}$-J model, the phase of the hopping parameter is found to be relatively small, especially at high doping levels. It makes the effective flux become smaller than the original Hamiltonian, causing the superconducting gap to become less dependent on $\phi$ as shown in Fig.\ref{Fig4}(c)-(d).  

\emph{Nematic nodal superconductor from strain}--So far we have only observed gapped chiral superconductor respecting $C_3$ symmetry. One may expect that a nematic superconductor with mixing between $l=1$ and $l=-1$ angular momentum channels is a competing state. In the following we show that a small strain can easily favor a nematic nodal superconductor.  We consider the effects of strain by multiplying a factor $(1-\alpha)$ on the hopping parameter $te^{i\phi^s_{ij}}$ on the bond along the $x$-axis, i.e., $t_{i\pm\mathbf{a_1},i}$, where $\alpha$ implies the stress intensity. Since $J_{ij}=\frac{t_{ij}^2}{U}$, the $J$ term needs to be multiplied by a factor $(1-\alpha)^2$ along the $x$-axis. When $\alpha=1$, the Hamiltonian is reduced to the one on the square lattice, where  the pairing symmetry is known to be $d_{x^2-y^2}$.  Therefore, the time reversal symmetry and mirror reflection symmetry can be restored by sufficently large strain.  During the calculation, the pairing parameter $\Delta_{i\pm\mathbf{a_1},i}$ is found to decay very 
fast when $\alpha$ grows. It shows that the $d_{x^2-y^2}$ superconductor in the square lattice can be reached by small $\alpha$. We use $1-\braket{\mathcal{T}}$ and $1-\braket{\mathcal{M}_y}$ to label the symmetry of the ground states. Here $\braket{\mathcal{T}}$ is defined by the inner product of the ground states of $H$ and $\mathcal{T}H\mathcal{T}^{-1}$, and similar for $\mathcal{M}_y$. Since the Hamiltonian has a global U(1) symmetry, $\braket{\mathcal{T}}$ and $\braket{\mathcal{M}_y}$ will change as the global phase $\theta$ changes. Therefore, we need to choose the appropriate $\theta$ to maximize the expected value to represent the symmetry faithfully. As shown in Fig.\ref{Fig6}, the time reversal symmetry and mirror reflection symmetry can be restored by a small strain.  In Fig.\ref{Fig6}(d), we show that there are four nodes in the resulting superconductor phase under strain.  This calculation demonstrates that  a nodal nematic superconductor is indeed a strong competing state and can be favored by a small strain. It is interesting to also study the possibility that the $C_3$ symmetry is spontaneously broken by additional terms.

\begin{figure}[h]
    \centering
    \includegraphics[width=\linewidth]{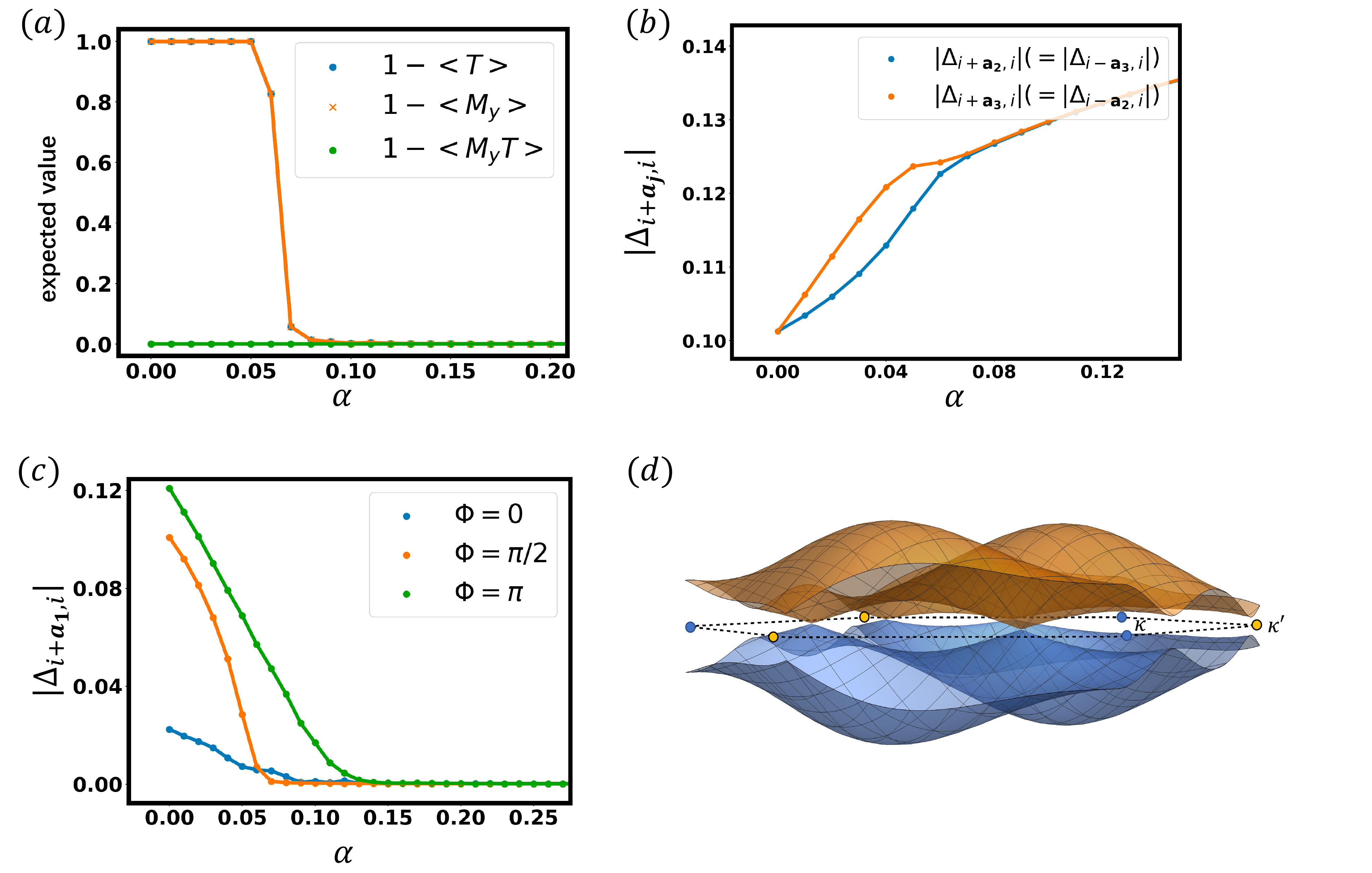}
    \caption{(color online). $t/J=3$, $n=0.85$. (a) The expected value of time reversal operator, mirror reflection operator and mirror time reversal operator as $\alpha$ changes for flux $\Phi=\frac{\pi}{2}$. (b) Dependence of $|\Delta_{i+\mathbf{a_2},i}|(=|\Delta_{i-\mathbf{a_3},i}|)$ and $|\Delta_{i+\mathbf{a_3},i}|(=|\Delta_{i-\mathbf{a_2},i}|)$ on $\alpha$ for $\Phi=\frac{\pi}{2}$ with the chirality fixed. (c) Dependence of $|\Delta_{i+\mathbf{a_1},i}|$ on $\alpha$ for flux $\Phi=0,\frac{\pi}{2},\pi$. (d) The dispersion plot for $\Phi=\frac{\pi}{2},\alpha=0.15$, the gap closes at $4$ nodal points in the first Brillouin zone.}
    \label{Fig5}
\end{figure}

\emph{The effect of three-site hopping term}-- In addition to Eq.\ref{eqtJ}, perturbation theory also gives a three-site hopping term \cite{PhysRevB.52.629}, it can be written as:
\begin{equation}
\begin{split}
    H_{t_3}&=-t_3\sum_{\braket{ijk},s}P\Big(n_{j,-s}c^\dagger_{k,s}c_{i,s}e^{i2\phi^s_{kj}}\\
    &-c^\dagger_{k,s}c^\dagger_{j,-s}c_{j,s}c_{i,-s}+\text{H.c.}\Big)P,
\end{split}
\end{equation}
where $t_3=\frac{J}{4}$. Here we suppose $t_3$ can be changed independently in order to discuss this term's effect. By applying the slave boson approach mentioned above, we can read off this term as:
\begin{equation}\label{eqsbt3}
 \begin{split}
    H_{t_3}&=-t_3\sum_{\braket{ijk},s}\Big(b_kb^\dagger_if^\dagger_{j,-s}f_{j,-s}f^\dagger_{k,s}f_{i,s}e^{i2\phi^s_{kj}}\\
    &-b_kb^\dagger_if^\dagger_{k,s}f^\dagger_{j,-s}f_{j,s}f_{i,-s}+\text{H.c.}\Big),
\end{split}
\end{equation}
where $b_kb^\dagger_i\approx\braket{b_kb^\dagger_i}=x$ is the doping level, so this term's effect increases as the doping level increases. Indeed, Eq.\ref{eqsbt3} can be decoupled in the pairing channel \cite{PhysRevB.52.629} as:
\begin{equation}
    H_{t_3}=xt_3\sum_{\braket{ij}}\left[\Delta_{ij}(\Delta^*_{ji}e^{i2\phi^\uparrow_{ji}}+\Delta^*_{ij})+\text{H.c.}\right]
\end{equation}
 So we can expect that a positive value of $t_3$ can increase the total energy, leading to a decrease in $|\Delta_{ij}|$ by solving the self-consistent equation. The mean field calculation validates our expectation as shown in Fig.\ref{Fig6}(a). This suggests that the $n>1$ side would have stronger superconductor than the $n<1$ side even for the particle-hole symmetric point $\Phi=\frac{\pi}{2}$.

 \begin{figure}[h]
    \centering
    \includegraphics[width=\linewidth]{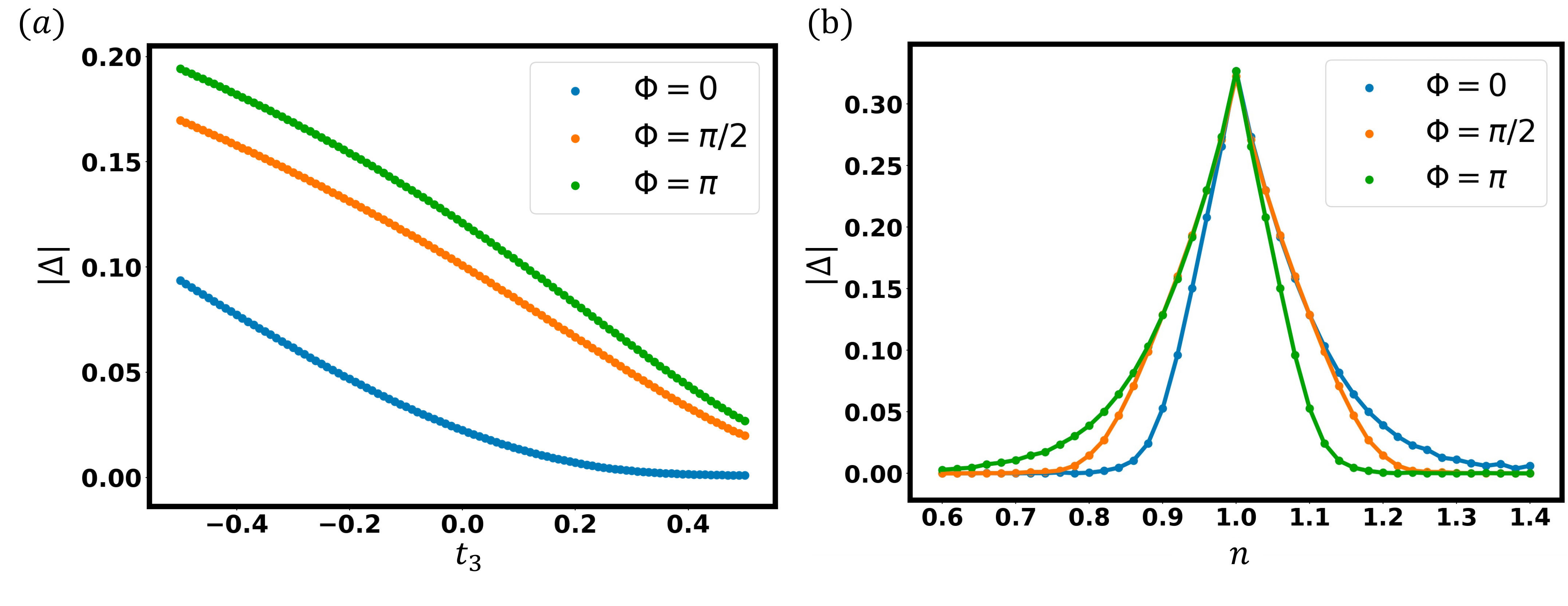}
    \caption{(color online). $t/J=3$. (a) Dependence of $|\Delta|$ on $t_3$ for $n=0.85$. (b) Dependence of $|\Delta|$ on $n$ for $t_3=0.25J$.}
    \label{Fig6}
\end{figure}
 
\emph{Summary}-- In conclusion, we use a slave boson method to investigate a t-J model with valley contrasting flux in the context of twisted TMD homobilayer. We show that the superconducting order parameter is a mixture of $p-ip$ and $d+id$  pairing when $\phi\ne0$.  We notice two topological phase transitions with jump of Chern numbers by tuning the valley contrasting flux $\Phi$, which is controlled by the vertical displacement field in the TMD homo-bilayer.  The pairing strength correlates with the density of states (DoS), suggesting that we should search for superconductor near the van Hove singularity. Finally, we find that a small strain can favor a nematic nodal superconductor similar to the $d_{x^2-y^2}$ pairing on the square lattice.

\emph{Acknowledgement} This work is supported by a startup fund from the Johns Hopkins University.

\bibliography{bibfile}

\appendix
\onecolumngrid
\section{Topological transitions}\label{appendixA}
\begin{figure}[h]
    \centering
    \includegraphics[width=\linewidth]{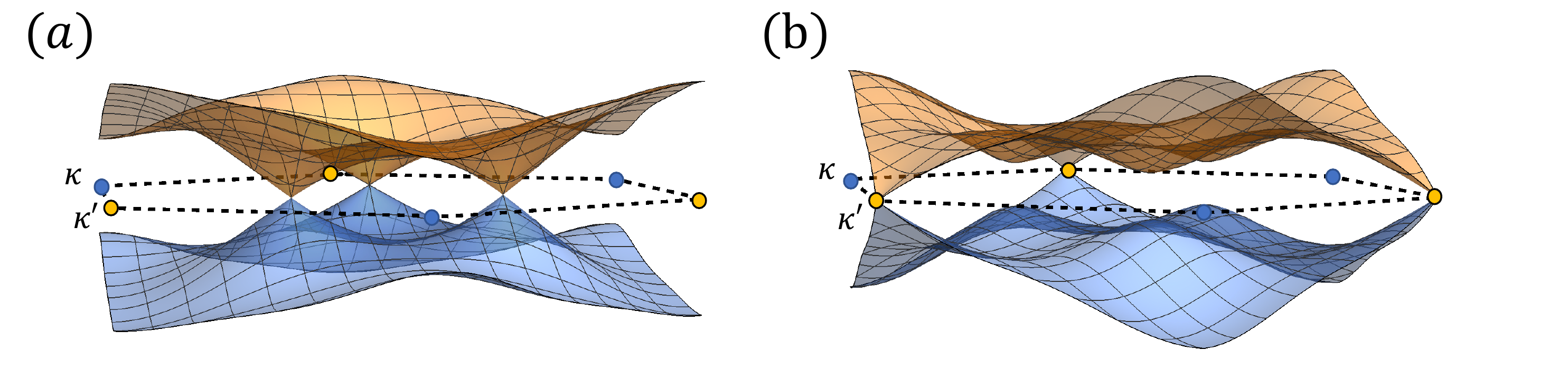}
    \caption{(color online). (a) Illustration of the three Dirac nodes inside the first Brillouin zone. (b) Illustration of the Dirac node at $\kappa'$.}
    \label{Fig_appendix_1}
\end{figure}

In the main text, we observe that Chern number depends on $\Phi$ and $n$. There are two kinds of topological transitions, which corresponds to one Dirac node and three Dirac nodes respectively. The sketch of the dispersion plot for two different situations is shown in Fig.\ref{Fig_appendix_1}.

\section{Stability of the Dirac nodes}

\begin{figure}[h]
    \centering
    \includegraphics[width=\linewidth]{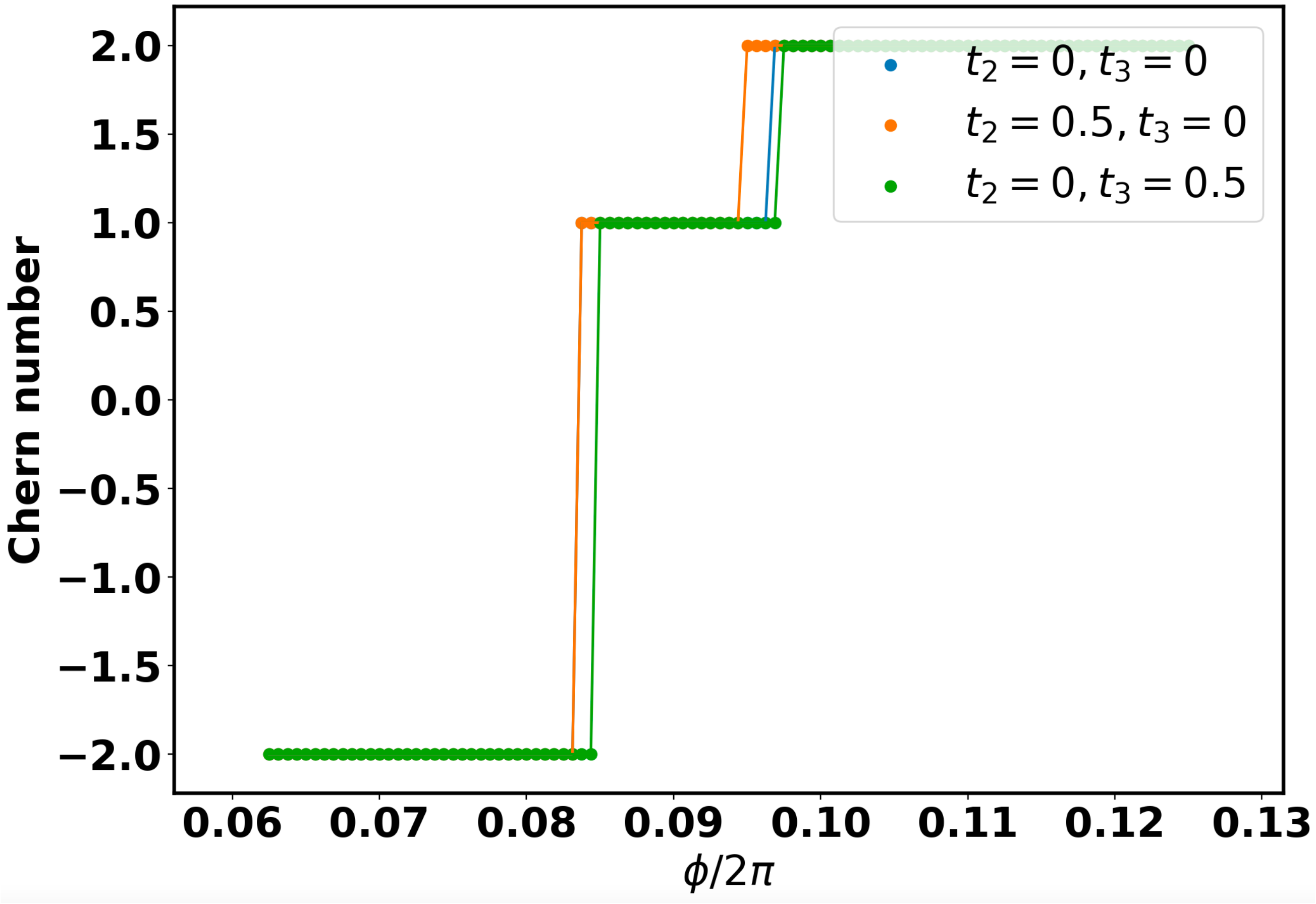}
    \caption{(color online). Dependence of the Chern number on $\phi$ at $n=0.85$.}
    \label{Fig_appendix_2}
\end{figure}

In the main text, we show that the Chern number has two kinds of transitions while variating the parameters. Here we demonstrate that the transitions will remain the same type under perturbations respecting $C_3$ symmetry. This fact can be verified by calculating the number of the Dirac nodes at the transition point. The Dirac node corresponds to the point in the first Brillouin zone where $\Delta(\mathbf{k})$ and $\epsilon(\mathbf{k})$ vanish simultaneously. In our system, the $C_3$ symmetry in the Hamiltonian requires that $\Delta(C_3\mathbf{k})=\Delta(\mathbf{k})$ and $\epsilon(C_3\mathbf{k})=\epsilon(\mathbf{k})$. Therefore, the Dirac node in the transition points that the Chern number only change by $1$ locates at $\Gamma$, $\kappa$ or $\kappa^\prime$. The superconducting order parameter in $k$-space can be written as:
\begin{equation}
    \Delta(\mathbf{k})=2|\Delta|\sum_{j=1,2,3}e^{i\frac{2\pi}{3}(j-1)}\cos(\mathbf{k}\cdot\mathbf{a_j}-\phi_{\textbf{SC}}),
\end{equation}
here $\phi_{\textbf{SC}}$ is defined by the relation $\Delta_{i+\mathbf{a_j},i}=e^{-i2\phi_{\textbf{SC}}}\Delta_{i-\mathbf{a_j},i}$. The expression around $\Gamma$, $\kappa$ or $\kappa^\prime$ can be expanded as:
\begin{equation}
    \Delta(\mathbf{k})=3|\Delta|\beta(k_x+ik_y),
\end{equation}
where $\beta=\sin\phi_{\text{SC}},\sin(\phi_\text{SC}+\frac{2\pi}{3}),\sin(\phi_\text{SC}-\frac{2\pi}{3})$ corresponds to $\Gamma$, $\kappa$, $\kappa^\prime$ respectively. Any rotation around $z$-axis is equivalent to a $U(1)$ gauge transform. The solutions of $\Delta(\mathbf{k})=0$ can only be shifted by adding a constant, which breaks the $z-$rotation symmetry. Therefore, we can conclude that the existence of the Dirac node at $\Gamma$, $\kappa$ or $\kappa^\prime$ is robust under perturbations respecting $C_3$ symmetry. This can be illustrated by considering the three-site term \ref{eqsbt3} and next-nearest-neighbor (NNN) hopping:
\begin{equation}
    H_{t_2}=-t_2\sum_{\braket{\braket{ij}},s}P(c^\dagger_{i,s}c_{j,s}+\text{H.c.})P,
\end{equation}
where $\braket{\braket{ij}}$ represents NNN sites. As shown in Fig.\ref{Fig_appendix_2}, the transition of the Chern number remains the same pattern.
\section{Symmetry transformation on the order parameters}
We note that the order parameters in Eq.\ref{MF} has $12$ degrees of freedom, in which $6$ corresponds to the pairing in six bonds, they are $\Delta_{i\pm\mathbf{a_j},i}, j=1,2,3$, where $\mathbf{a_1}=(1,0),\mathbf{a_2}=(-\frac{1}{2},\frac{\sqrt{3}}{2}),\mathbf{a_3}=(-\frac{1}{2},-\frac{\sqrt{3}}{2})$ and $\mathbf{a_{j+3}}=\mathbf{a_j}$.  The remaining $6=3+3$ corresponds to the hoping parameters in spin up and down channels, they are $\chi_{i+\mathbf{a_j},i,\uparrow}$ and $\chi_{i+\mathbf{a_j},i,\downarrow}, j=1,2,3$. Under $C_6$ rotation, the order parameters transform as $\Delta_{i+\mathbf{a_j},i}\rightarrow-\Delta_{i+\mathbf{a_{j-1}},i}$, $\chi_{i+\mathbf{a_j},i,s}\rightarrow\chi_{i-\mathbf{a_{j-1}},i,-s}$. Under time reversal transformation, the order parameters transform as $\Delta_{i\pm\mathbf{a_j},i}\rightarrow-\Delta_{i\mp\mathbf{a_j},i}^*$ and $\chi_{i\pm\mathbf{a_j},i,s}\rightarrow\chi^*_{i\pm\mathbf{a_j},i,-s}$. Under mirror reflection transformation about the $x$-axis, the order parameters transform as $\Delta_{i\pm\mathbf{a_j},i}\rightarrow\Delta_{i\pm\mathbf{a_{5-j}},i}$ and $\chi_{i\pm\mathbf{a_j},i,s}\rightarrow\chi_{i\pm\mathbf{a_{5-j}},i,s}$. Under mirror reflection transformation about the $y$-axis, the order parameters transform as $\Delta_{i\pm\mathbf{a_j},i}\rightarrow-\Delta_{i\pm\mathbf{a_{5-j}},i}$ and $\chi_{i\pm\mathbf{a_j},i,s}\rightarrow\chi_{i\mp\mathbf{a_{5-j}},i,-s}$.

From the above analysis, one can verify that under mirror time reversal transformation about the $x$-axis, the order parameters transform as $\Delta_{i\pm\mathbf{a_j},i}\rightarrow-\Delta^*_{i\mp\mathbf{a_{5-j}},i}$ and $\chi_{i\pm\mathbf{a_j},i,s}\rightarrow\chi^*_{i\pm\mathbf{a_{5-j}},i,-s}$. Under mirror time reversal transformation about the $y$-axis, the order parameters transform as $\Delta_{i\pm\mathbf{a_j},i}\rightarrow\Delta^*_{i\mp\mathbf{a_{5-j}},i}$ and $\chi_{i\pm\mathbf{a_j},i,s}\rightarrow\chi^*_{i\mp\mathbf{a_{5-j}},i,s}$. The mean field results give that the order parameters are gauge equivalent to the ones before the symmetry transformation.

\end{document}